\documentclass[]{tMOP2e}

\begin{document}

\doi{10.1080/0950034YYxxxxxxxx}
 \issn{1362-3044}
\issnp{0950-0340} \jvol{00} \jnum{00} \jyear{2008} \jmonth{10 January}

\markboth{Kadic, Dupont, Guenneau and Enoch}{Journal of Modern Optics}

\title{Controlling surface plasmon polaritons in transformed coordinates}

\author{Muamer Kadic, Guillaume Dupont, Sebastien Guenneau, Stefan Enoch \thanks{$^\ast$Corresponding author. Email: sebastien.guenneau@fresnel.fr}
\\{\em{Institut Fresnel, CNRS, Aix-Marseille
Universit\'e,\\Campus universitaire de Saint-J\'er\^ome,
 13013 Marseille, France}}.
\\\vspace{6pt}\received{v3.3 released October 2010} }

\maketitle

\begin{abstract}
Transformational optics allow for a markedly enhanced control of the electromagnetic wave trajectories
within metamaterials with interesting applications ranging from perfect lenses to invisibility cloaks, carpets, concentrators
and rotators.
Here, we present a review of curved anisotropic heterogeneous meta-surfaces designed using the tool of transformational plasmonics, in order to achieve a similar control for surface plasmon polaritons in cylindrical and conical carpets, as well as
cylindrical cloaks, concentrators and rotators of a non-convex cross-section. Finally, we provide an asymptotic form of the geometric potential for surface plasmon polaritons on such surfaces in the limit of small curvature.   
\end{abstract}

\bigskip

\section{Introduction to transformational optics}
It has been known since the advent of mathematical optics \cite{luneburg}
that lenses with certain curvature and/or a spatially varying
refractive index allow for focussed images. Indeed, the simplest example of a convergent lens is
a curved piece of silica which bends light according to the Snell-Descartes
laws of refraction at curved air/glass interface. However, such a lens
is necessarily of a convex shape. On the contrary, a medium with a continuously changing refractive index can act as a flat convergent
lens: Gradient-index optics is indeed the branch of optics covering optical effects produced by a gradual variation of the refractive index of a material with the famous paradigm of the Maxwell fisheye, first published by James Clerck
Maxwell in 1854 under the pseudonym of Charles Anthony Swainson \cite{max1}, but however still of current interest \cite{ulf1,jmo2010}.
A continuously refracting index can also create optical illusions known as mirages: in the desert, these are due to the
variation of refractive index between the hot, less dense air at the surface of the road, and the denser cool air above it.
The variation in temperature (and thus density) of the air actually causes a gradient in its refractive index, causing it to increase with height. This index gradient causes anomalous refraction of light rays (at a shallow angle to the road) from the sky, bending them into eye of the viewer, with their apparent location being the road's surface. Similar effects occur at the surface of the sea, and are known as Fata Morgana, after the name of the fairy in the Arthurian legend who lifted castles
over lakes. Inverted optics include perfect lenses via negative refraction \cite{pendry_prl00,sar3,pendry_jpc03,guenneau_ol05}, that can also be designed using transformational optics by
mapping the image plane onto the source plane, which causes a negative (possibly spatially varying) refractive index \cite{ulf1,sgsar09}. Maybe the most stricking example of an optical illusion is the invisibility cloak independently proposed
by Pendry, Schurig and Smith, and Leonhardt
\cite{leonhardt06,pendry06b}, which can steer light around an object which is then invisible to an external observer (however, light emanating from this object is itself confined within the invisibility cloak!). The underlying mechanism
of such cloaks relies upon a mirage effect, since a source located inside the coating seems to radiate from
a shifted location \cite{zolla}.

This coating actually consists of a meta-material whose physical properties are deduced from a coordinate
transformation in the Maxwell system, and this viewpoint allows for a refreshed analysis of geodesics of
light in complex media, since the Maxwell equations retain their form under geometric changes and the
only thing to do in the transformed coordinates is to replace the original medium (often isotropic homogeneous)
by an anisotropic heterogeneous medium \cite{ward,nicolet} (a meta-material).
Such geometric transforms are very useful in computational electromagnetism to model unbounded
\cite{compel2001} and twisted \cite{ejp2004} domains in the context of optical fibres.
However, the physical interest in transformational optics has been fuelled by potential applications
in the design of meta-materials, while the electrical engineering community has used geometric transforms
as a mean to simplify numerical models involving large or thin domains for the past thirty years. 

Coming back to the invisibility cloak, the anisotropy and the heterogeneity of the parameters of its coat work
as a deformation of the optical space around the object. The first
experimental validation \cite{pendry06b} of these theoretical
considerations was given, a few months after the proposals of Pendry et al., and Leonhardt
were released in the press. A copper cylinder was indeed made invisible to an incident plane wave at
$8.5$ GHz as predicted by the numerical simulations. This markedly enhance our capabilities to
manipulate light, even in the extreme near field limit \cite{zolla}.
However, such cloaks suffer from an inherent narrow bandwidth as
their transformation optics design leads to singular tensors on the
frontier of the invisibility region which is a consequence of
tearing apart the metric when one makes a hole in optical space
(obtained by blowing up a point \cite{greenleaf}).
Physically, light has to curve its trajectory around the hole (or 
'invisibility region'); Hence, to match the phase of a wave
propagating in homogeneous space, it must travel faster.
To remove the cloak's singularity, Xiang et al.
proposed to consider the blowup of a segment instead of a point
\cite{jiang}, but this cloak only works for certain directions. On
the other hand, Leonhardt and Tyc considered a stereographic
projection of a virtual hyper-sphere in a four dimensional space
\cite{tyc}, which bears some resemblance with the construction of a
Maxwell fisheye.

As an alternative to non-singular cloaking Li and Pendry proposed a
one-to-one geometric transform from a flat to a curved ground: their
invisibility carpet \cite{lipendry} is by essence non singular and
thus broadband. This proposal led to a rapid experimental progress in the
construction of two-dimensional carpets approaching optical frequencies
\cite{smithcarpet,zhangcarpet1,gabrielli-carpet}.

Another way to make cloaks broadband is to approximate
their parameters using a homogenization approach, which leads to
nearly ideal cloaking \cite{naturecloak,farhat,kurylev}, as it does
not rely upon locally resonant elements.
In 2008, some of us \cite{farhat} demonstrated broadband
cloaking of transverse electric waves using a micro-structured metallic
cloak in the micro-wave regime.
This naturally prompts the question of whether, at optical frequencies, an object lying onto a metal film
could be cloaked from a propagating Surface Plasmon Polaritons (SPPs).

\section{From transformational optics to plasmonics}

The less than usual physics of the transmission of light through holes small
compared with the wavelength is has given rise to a vast amount of literature following
the 1998 paper by Ebbesen et al. \cite{ebbesen}, but some
heralding earlier work combining both theory and experiment is less well
known \cite{ross80}.
Martin-Moreno and Garcia-Vidal further showed in 2004 that one can
manipulate surface plasmon ad libitum via homogenization of
structured surfaces \cite{science2004}. In the same vein, pioneering
approaches to invisibility relying upon plasmonic metamaterials have
already led to fascinating results
\cite{milton2,engheta,javier,baumeier}. These include plasmonic
shells with a suitable out-of-phase polarizability in order to
compensate the scattering from the knowledge of the electromagnetic
parameters of the object to be hidden, and external cloaking, whereby a
plasmonic resonance cancels the external field at the location of a
set of electric dipoles. Recently, Baumeier et al. have demonstrated
theoretically and experimentally that it is possible to reduce
significantly the scattering of an object by a SPP, when it is surrounded by two concentric rings of point
scatterers \cite{baumeier}.

In the present paper, we would like to extend the control of electromagnetic fields using tools of geometric transforms to the area of surface plasmons.
To do this, we adopt the general form for the time-harmonic electric and magnetic field, $\bm{E}$ and $\bm{H}$
in $exp(j\omega t)$.
\begin{center}
 \begin{equation}
  \left\{ \begin{array}{ccl}
           \bm{E}(x,y,z,t) & = & \bm{E}(x,y,z)e^{j\omega t}\\
          \bm{H}(x,y,z,t) & = & \bm{H}(x,y,z)e^{j\omega t}
          \end{array} \right.
 \end{equation}
\end{center}
where $\omega$ denotes the angular frequency and $t$ is the time variable.
We then consider the time-harmonic Maxwell's equations in the initial space coordinates (x,y,z):
\begin{center}
 \begin{equation}
  \left\{ \begin{array}{ccl}
           \nabla \times \bm{E} & = & -j\omega \underline{\underline{\mu}}\bm{H}\\
         \nabla \times \bm{H} & = & j\omega \underline{\underline{\varepsilon}}\bm{E}\\
           \nabla .(\underline{\underline{\varepsilon}}\bm{E}) & = & 0\\
        \nabla .(\underline{\underline{\mu}}\bm{H}) & = & 0
       \end{array} \right. \label{maxwsyst1}
\end{equation}
\end{center}
where the tensors of permittivity and permeability $\underline{\underline{\varepsilon}}$ and $\underline{\underline{\mu}}$ describe the
original (possibly heterogeneous anisotropic) medium.

In the transformed coordinate system (u,v,w), the structure of the Maxwell equations is preserved:
\begin{center}
 \begin{equation}
  \left\{ \begin{array}{ccl}
           \nabla' \times \bm{E} & = & -j\omega \underline{\underline{\mu'}}\bm{H}\\
         \nabla' \times \bm{H} & = & j\omega \underline{\underline{\varepsilon'}}\bm{E}\\
           \nabla' .(\underline{\underline{\varepsilon'}}\bm{E}) & = & 0\\
        \nabla' .(\underline{\underline{\mu'}}\bm{H}) & = & 0
       \end{array} \right. \label{maxwsyst1}
\end{equation}
\end{center}
but we note that the space derivatives are now taken with respect to the transformed coordinates (hence, the transformed gradient $\nabla'$), and obviously the
permittivity and permeability tensors $\underline{\underline{\varepsilon'}}$ and $\underline{\underline{\mu'}}$ should have new expressions.


An elegant way to identify the tensors is to consider the Jacobian matrix $\bm{J}$ associated with the change of co-ordinate system:

\begin{center}
 \begin{equation}
  \left\{ \begin{array}{ccl}
          dx & = & \dfrac{\partial x}{\partial u}du + \dfrac{\partial x}{\partial v}dv + \dfrac{\partial x}{\partial w}dw\\
        dy & = & \dfrac{\partial y}{\partial u}du + \dfrac{\partial y}{\partial v}dv + \dfrac{\partial y}{\partial w}dw\\
        dz & = & \dfrac{\partial z}{\partial u}du + \dfrac{\partial z}{\partial v}dv + \dfrac{\partial z}{\partial w}dw
\end{array} \right. \quad \Longleftrightarrow \quad
  \left( \begin{array}{c}
          dx\\
        dy\\
        dz
         \end{array} \right) = \bm{J}
  \left( \begin{array}{c}
          du\\
        dv\\
        dw
         \end{array} \right)
 \end{equation}
\end{center}

It follows that \cite{nicolet} the transformation rule for the tensors is:
$(\underline{\underline{\varepsilon}},\underline{\underline{\mu}})\rightarrow
(\underline{\underline{\varepsilon'}},\underline{\underline{\mu'}})$ with:
\begin{center}
 \begin{equation}
  \left\{ \begin{array}{ccl}
\underline{\underline{\varepsilon'}}(u,v,w) & = & \bm{J}^{-1} \underline{\underline{\varepsilon}}(x,y,z)\bm{J}^{-T}{\rm det}(\bm{J})\\
\underline{\underline{\mu'}}(u,v,w) & = & \bm{J}^{-1} \underline{\underline{\mu}}(x,y,z)\bm{J}^{-T}{\rm det}(\bm{J})\\
\end{array}\right.
 \end{equation}
\end{center}
where ${\rm det}(\bm{J})$ is the determinent of the Jacobian matrix.

When the original permittivity and permeability matrices are scalar,
their transformed counterparts are given by:
\begin{equation}
{\underline{\underline{\varepsilon'}}}=\varepsilon {\bf T}^{-1}
\; , \;
{\underline{\underline{\mu'}}}=\mu {\bf T}^{-1}
\; , \;
\hbox{ where } {\bf T}=\frac{{\bf J}^T {\bf J}}{\rm{det}({\bf J})}
\end{equation}


We now wish to apply this general rule to the case of a transverse magnetic ($p$-polarized) SPP propagating in the positive $x$ direction at the
interface $z=0$ between metal ($z<0$) and air ($z>0$):
\begin{equation}
\left\{
\begin{array}{ll}
{\bf H}_2 &= (0,H_{y2},0) \exp \{\i(k_{x2}x-\omega t)-k_{z2}z\} \; , z>0 \; ,\\
{\bf H}_1 &= (0,H_{y1},0) \exp \{\i(k_{x1}x-\omega t)+k_{z1}z \} \;
, z<0 \; ,
\end{array}
\right. \label{sppview}
\end{equation}

where $c$ is the speed of light in vacuum and
$\varepsilon_2=1-\frac{\omega_p^2}{\omega^2+i\gamma\omega}$ has the
usual Drude form in the metal($z<0$), for which $\omega_p$ is the
plasma frequency ($2175$ THz) of the {\it free electron gas} and
$\gamma$ is a characteristic collision frequency of about $4.35$ THz
\cite{palik}. Moreover, the upper medium is described by spatially varying
tensors of permittivity $\underline{\underline{\varepsilon'}}$ and permeability
$\underline{\underline{\mu'}}$ which for simplicity are assumed to be represented
in a diagonal basis i.e. $\underline{\underline{\varepsilon'}}={\rm diag}(\varepsilon_{xx2},\varepsilon_{yy2},\varepsilon_{zz2})$
and $\underline{\underline{\mu'}}={\rm diag}(\mu_{xx2},\mu_{yy2},\mu_{zz2})$.  

\noindent The dispersion relation for the surface polariton at such an anisotropic interface takes the following form \cite{kadic}
\begin{equation}
k_x=\frac{\omega}{c}\sqrt{\frac{\varepsilon_{zz2}\varepsilon_1
(\mu_{yy2}\varepsilon_1-\varepsilon_{xx2})}{\varepsilon_1^2-\varepsilon_{xx2}\varepsilon_{zz2}}}
\; . \label{spp1new}
\end{equation}
SPPs are bound to the interface, hence do not belong to the radiative spectrum (unlike leaky waves).

\section{Design of plasmonic paradigms}
In the present section we show numerically that one can control the SPP field using transformational plasmonics.
A review of transformations leading to arbitrary carpets, concentrators and rotators is presented.

\subsection{Three-dimensional carpets for electromagnetic and plasmonic fields}

So far, such a mathematical derivation of the SPP dispersion relation seems fairly abstract. However, we
now wish to analyse the interaction of this SPP with a specific anisotropic
heterogeneous structure, in the present case a three-dimensional invisibility carpet \cite{pra2010},
deduced from the following geometric transformation:

\begin{equation}
\left\{ \begin{array}{c c l}
            x' & = & x \\
	    y' & = & y \\
	    z' & = & \dfrac{z_2 - z_1}{z_2} z + z_1.\\
           \end{array} \right. 
\end{equation}

Importantly, we note that this material is not only heterogeneous anisotropic
but also magnetic, which seems a far technological reach. However,
these constraints can be further relaxed using some quasi-conformal
grids in the spirit of Li and Pendry's work for two-dimensional carpets \cite{lipendry}.

Here $z'$ is a stretched coordinate. It is easily seen
that this linear geometric transform maps the surface $z_0(x,y)$ of the
horizontal plan $z(x,y)=0$ onto the surface $z(x,y)=z_1(x,y)$, and it leaves the
surface $z(x,y)=z_2(x,y)$ unchanged.

The surfaces $z_1$ and
$z_2$ are assumed to be differentiable, and this ensures that the
carpet won't display any singularity on its inner boundary.

\noindent The symmetric tensors
$\underline{\underline{\varepsilon'}}$ and
$\underline{\underline{\mu'}}$ are fully described by five non
vanishing entries in a Cartesian basis:

\begin{figure}[h]
\begin{center}
\subfigure[Metal surface with a bump]{
\resizebox*{6cm}{!}{\includegraphics{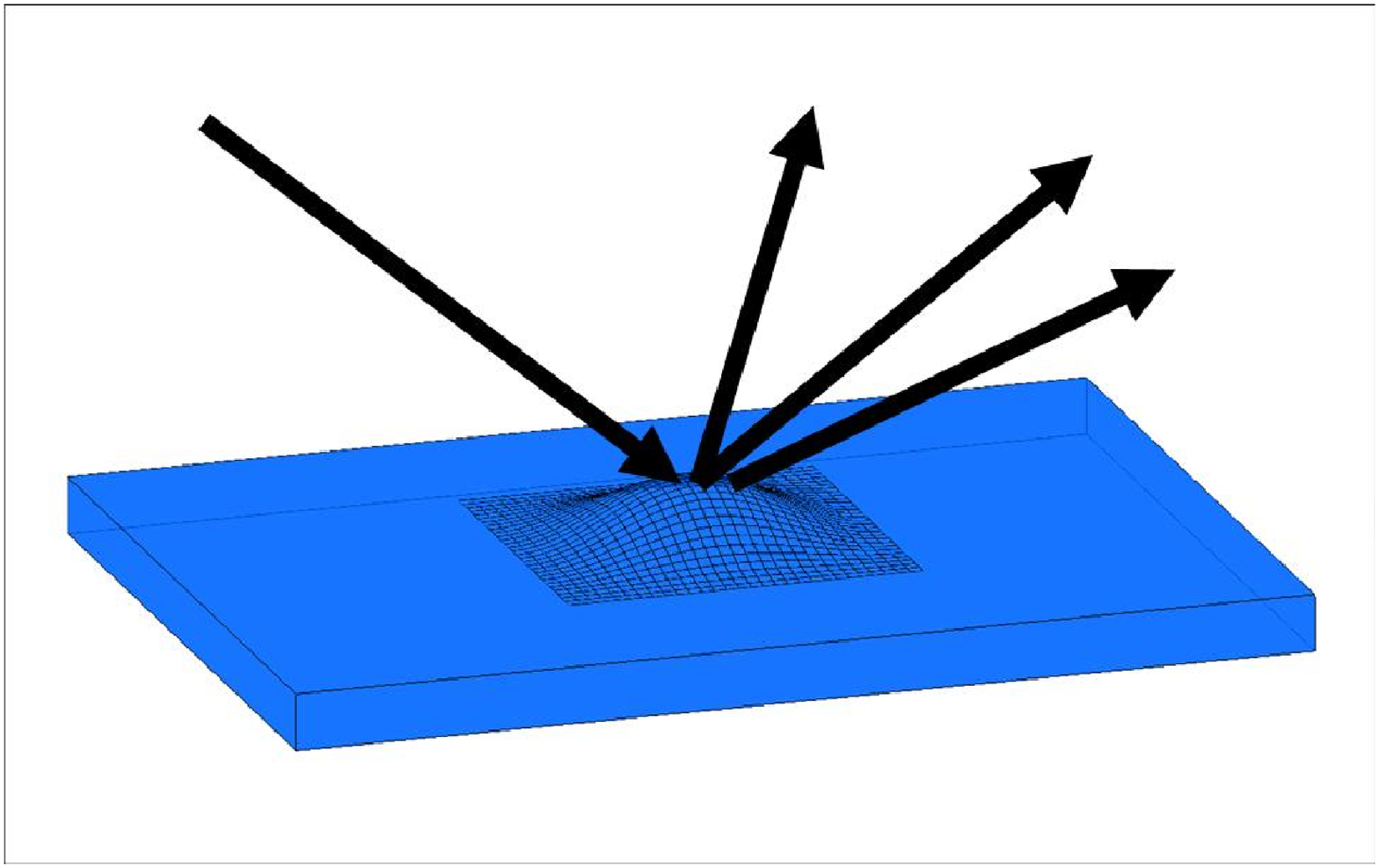}}}
\subfigure[Metal surface with a bump dressed with a transformed medium (invisibility carpet)]{
\resizebox*{6cm}{!}{\includegraphics{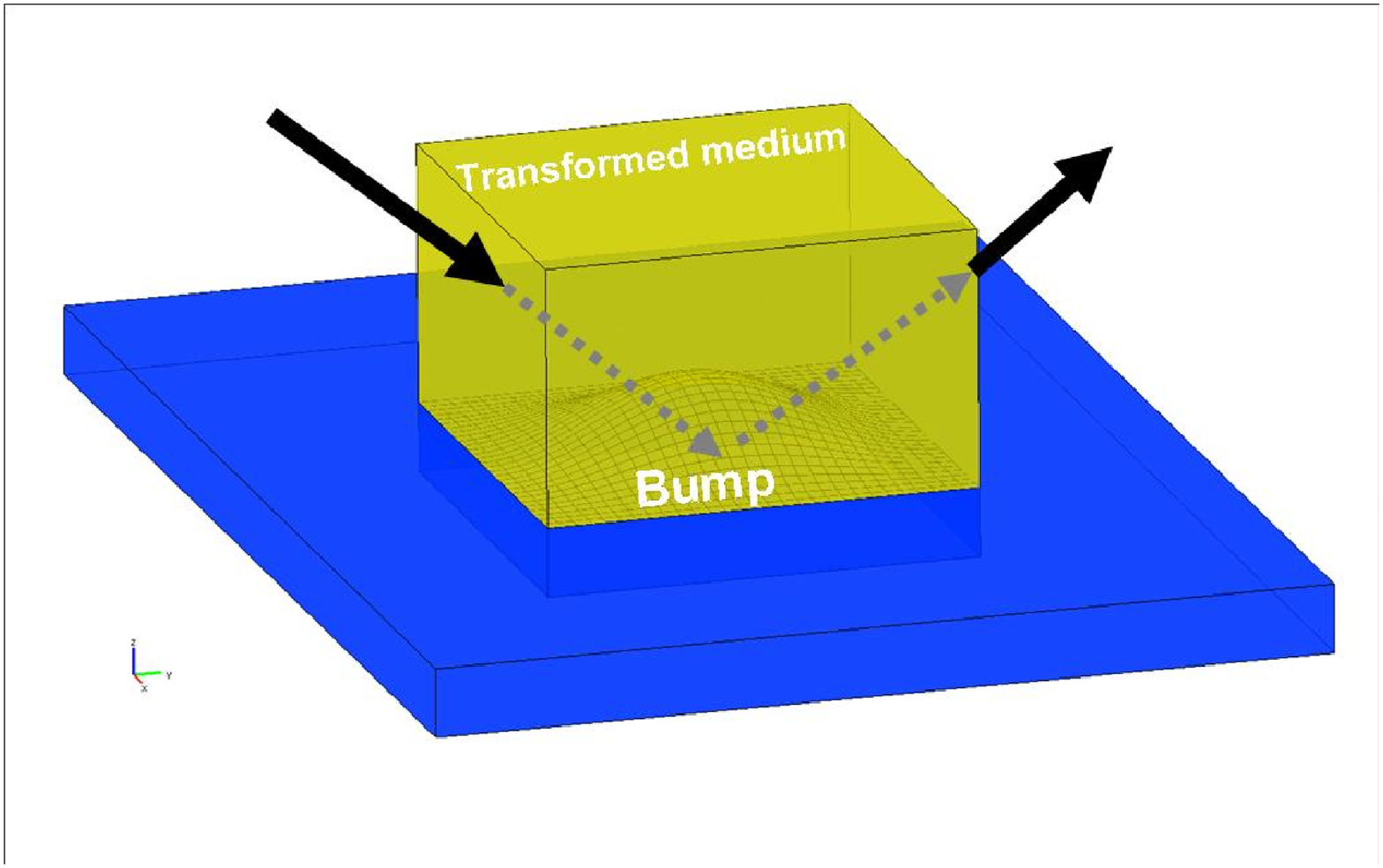}}}%
\caption{Principle of a three-dimensional invisibility carpet for electromagnetic fields: (a) Light incident upon a curved surface undergoes different orders of diffraction (shown as black arrows); (b)
 Light incident upon a curved surface with an invisibility crapet (yellow box) undergoes the same diffraction as light incident upon a flat metal surface.}
\end{center}
\label{fig1}
\end{figure}

\begin{figure}[h]
\begin{center}
\resizebox*{7cm}{!}{\includegraphics{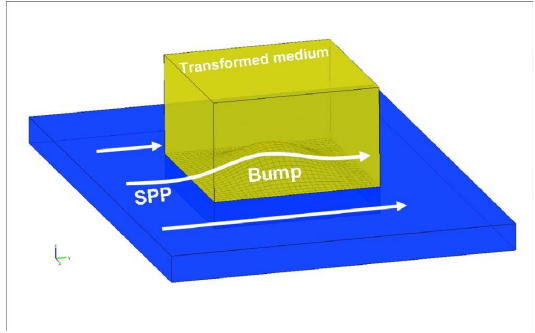}}
\caption{Principle of a three-dimensional invisibility carpet for plasmonic fields: A surface plasmon polariton incident from the left
(with trajectory shown as white arrows) on the metallic surface is smoothly bent in the metamaterial box (yellow box). It turns out that
the transformed media in the yellow box is the same as in Figure \ref{fig1}.}
\end{center}
\label{fig2}
\end{figure}


\begin{equation}
{\bf J}_{zz'} = \left( \begin{array}{c c c}
            1 & 0 & 0 \\
	    0 & 1 & 0 \\
	    \dfrac{\partial z}{\partial x'} & \dfrac{\partial z}{\partial y'} & \alpha^{-1} \\
           \end{array} \right) \quad \Longrightarrow
\quad \underline{\underline{\varepsilon'}}=\underline{\underline{\mu'}}=  \left( \begin{array}{c c c}
            \alpha^{-1} & 0 & -\dfrac{\partial z}{\partial x'} \\
	    0 & \alpha^{-1} & -\dfrac{\partial z}{\partial y'} \\
	    -\dfrac{\partial z}{\partial x'} & -\dfrac{\partial z}{\partial y'} &
		  \alpha \left( 1+ \left( \dfrac{\partial z}{\partial x'} \right)^2 +\left( \dfrac{\partial z}{\partial y'} \right)^2 \right)\\
           \end{array} \right)
\end{equation}
where $\alpha=(z_2-z_1)/z_2$ and with $\mathbf{J}$ the Jacobian matrix of the transformation. Furthermore, the derivatives of $z$ with respect to $x'$ and $y'$ are given by:
\begin{equation}
\dfrac{\partial z}{\partial x'} = z_2 \dfrac{z' - z_2}{(z_2 - z_1)^2} \dfrac{\partial z_1}{\partial x'} +
	z_1 \dfrac{z_1 - z'}{(z_2 - z_1)^2} \dfrac{\partial z_2}{\partial x'} \quad ; \quad
\dfrac{\partial z}{\partial y'} = z_2 \dfrac{z' - z_2}{(z_2 - z_1)^2} \dfrac{\partial z_1}{\partial y'} +
	z_1 \dfrac{z_1 - z'}{(z_2 - z_1)^2} \dfrac{\partial z_2}{\partial y'}
\end{equation}
The purpose of the next section is to show that such carpets work equally well for electromagnetic and plasmonic fields, as is intuitively the case when looking at the limit of ray
optics, see Figures \ref{fig1} and \ref{fig2}: This is due to the fact that the transformed medium is valid for any field solution of the Maxwell's equations (which is obviously the
case for SPPs).

\subsection{Gaussian shaped bump:}

\begin{figure}[h]
\begin{center}
\resizebox*{10cm}{!}{\includegraphics{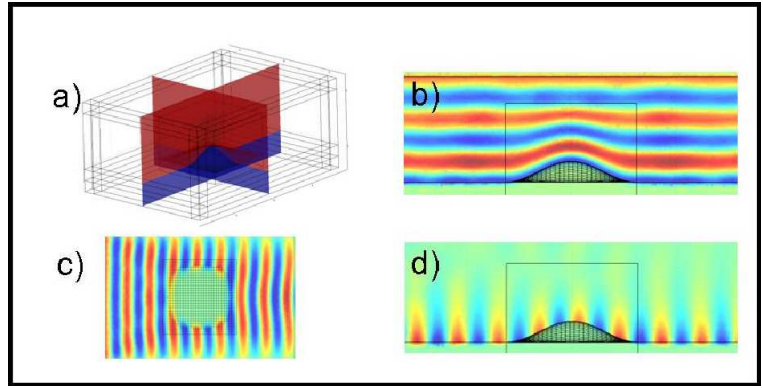}}
\caption{Plane wave incident from the top (z axis):
(a) Magnitude of the magnetic field; (b) Phase represention of the magnetic field. SPP launched from the left: (c) Slice of magnetic field along (x,y) i.e. top-view; (d) Slice of magnetic field along
(x,z) i.e. side view.}
\end{center}
\label{fig3a}
\end{figure}

We now wish to apply the recipe for the design of 3D carpets to specific geometrical transformations in order to hide Gaussian and conical shaped bumps on a metallic surface.
We first present the projection of the conical surface on the flat surface, as this case is more intuitive. The geometric transform used in such a design is as follows:
\begin{equation}
z_1= h_0 \cos\left( \dfrac{\pi x}{l}\right)^2\cos\left( \dfrac{\pi y}{l}\right)^2 \text{with } 
h= 2.10^{-7} \;;\; l = 1.25 \; 10^{-6}\; z_2 = 7.5 \; 10^{-7}\\
\end{equation}

\begin{equation}
\dfrac{\partial z_1}{\partial x} = -2 h_o\cos\left( \dfrac{\pi x}{l}\right).cos\left( \dfrac{\pi y}{l}\right)^2 .
\sin\left( \dfrac{\pi x}{l}\right).\dfrac{\pi}{l}\\
\end{equation}

\begin{equation}
\dfrac{\partial z_1}{\partial y} = -2 h_o\cos\left( \dfrac{\pi y}{l}\right)\cos\left( \dfrac{\pi x}{l}\right)^2 .
\sin\left( \dfrac{\pi y}{l}\right).\dfrac{\pi}{l}\\
\end{equation}
We report in Figure \ref{fig3a} the finite element computations which exemplify the role played by the carpet in the control of the SPP wavefront.

\subsection{Conical shaped bump: A less than unusual carpet}
\begin{figure}[h]
\begin{center}
\resizebox*{10cm}{!}{\includegraphics{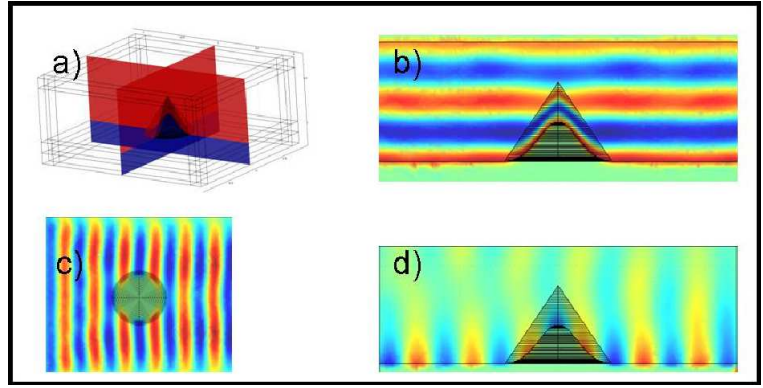}}
\caption{Plane wave incident from the top (z axis): (a) Magnitude of the magnetic field; (b) Phase represention of the magnetic field. SPP launched from the left: (c) Slice of magnetic field along (x,y) i.e. top-view; (d) Slice of magnetic field along
(x,z) i.e. side view.}
\end{center}
\label{fig3b}
\end{figure}
We now present the projection of the conical surface on the flat surface, as this case is less intuitive. For this, we consider two surfaces $z_1$ and $z_2$ which describe
the inner and outer boundaries of the carpet:

\begin{equation}
\begin{array}{ccc}
z_1 &= A\exp\left( -\dfrac{1}{2}\dfrac{x^2+y^2}{\sigma^2}\right)-b \quad \text{with}
\quad A = 0.205 \;;\; \sigma = 0.1 \;;\; b = 0.005 \\
z_2 &= h\left(1-\dfrac{\sqrt{x^2+y^2}}{L}\right) \quad \text{with}
\quad h = 0.4 \;;\; L = \sqrt{2\sigma^2 ln\left(\dfrac{A}{b}\right)} \\
\end{array}
\end{equation}

Explicit expressions for the partial derivatives are
\begin{equation}
\begin{array}{ccc}
\dfrac{\partial z_1}{\partial x} &= -\dfrac{A x}{\sigma^2} \exp\left( -\dfrac{1}{2}\dfrac{x^2+y^2}{\sigma^2}\right) \; , \;
\dfrac{\partial z_1}{\partial y} = -\dfrac{A y}{\sigma^2} \exp\left( -\dfrac{1}{2}\dfrac{x^2+y^2}{\sigma^2}\right) \; , \\
\dfrac{\partial z_2}{\partial x} &= -\dfrac{h x}{L \sqrt{x^2+y^2}} \; , \; \dfrac{\partial z_2}{\partial y} = -\dfrac{h y}{L \sqrt{x^2+y^2}} \; .
\end{array}
\end{equation}

We report the finite element computations in Figure \ref{fig3b}. We can see that control of electromagnetic and plasmonic fields is achieved
for both the smooth (Gaussian) and rough (conical shaped) bumps.

We emphasize that these numerical results are non-intuitive as the
scattering by a conical shaped surface is one of the hardest cases to handle. In order to see this, let us remind some classical result of
singularity theory for a sector (the two-dimensional counterpart of a conical shaped domain).

Let us consider the Laplace problem in a semi-infinite sector bounded by two edges $\Sigma_1$ and $\Sigma_2$:
\begin{equation}
-\Delta u = f \hbox{ dans } D \; , \; u=0 \hbox{ on } \Sigma_1 \hbox{ and } \Sigma_2 \; ,
\end{equation}
which we can express in polar coordinates:
\begin{equation}
\frac{\partial^2 u}{\partial r^2}  +
\frac{1}{r}\frac{\partial u}{\partial r} +
\frac{1}{r^2}\frac{\partial^2 u}{\partial \theta^2}
= 0 \hbox{ pour } 0\leq r<+\infty \; , \;  0\leq \theta\leq \omega \; , 
\end{equation}
and let us assume that Dirichlet boundary conditions on the edges $u(r,0)=u(r,\omega)=0$.

Using separation of variables, we find that
\begin{equation}
u(r,\theta)=\sum_{n=1}^{+\infty} \left( a_n r^{n\pi/\omega}
+ b_n r^{-n\pi/\omega}\right)
\sin(n\pi\theta/\omega) \; .
\label{fousect}
\end{equation}
   
It is well-known \cite{sashanatasha} that if $f$ is square integrable, then $u$ is of finite
energy,
that is
\begin{equation}
\int_{\hbox{sector}}{\mid \hbox{grad} u \mid}^2 dx
=\int_0^R \left(\int_0^\omega \left( {(\frac{\partial u}{\partial r})}^2 +
{(\frac{1}{r}\frac{\partial u}{\partial \theta})}^2 \right) d\theta\right) rdr < + \infty
\end{equation}

\noindent Therefore, we need to ensure that the following expressions are finite:
\begin{equation}
\frac{\omega}{2} \sum_n \int_0^R \left( {(a_n \frac{\pi}{\omega}r^{n\pi/\omega-1}
-b_n \frac{\pi}{\omega}r^{-n\pi/\omega-1})}^2 \right) \, rdr < + \infty
\end{equation}
\begin{equation}
\frac{\omega}{2} \sum_n \int_0^R 
\frac{n^2\pi^2}{\omega^2}
\left( {(a_n r^{n\pi/\omega}
+ b_n r^{-n\pi/\omega})}^2 \right) \, rdr < + \infty \; .
\end{equation}

\noindent We note that
$
\frac{\omega}{2} \sum_n \int_0^R b_n^2 {(\frac{\pi}{\omega}r^{-n\pi/\omega-1})}^2 \, rdr
$
is finite if and only if $-2n\pi/\omega-2+1>-1$. This condition is not met for $0<\omega<2\pi$.
Hence, all coefficients $b_n$ vanish in (\ref{fousect}).

\noindent The solution of the Dirichlet problem is thus given by:
\begin{equation}
u(r,\theta)=\sum_{n=0}^{+\infty} a_n r^{n\pi/\omega} \sin(\frac{n\pi\theta}{\omega}) \; .
\end{equation}

\noindent From this expression, we can see that the field $u$ is singular near the tip of the
sector i.e. when $r$ is close to zero. Similar expresions hold for a conical shaped region in
the three-dimensional case. However, when the sector/conical shaped region is filled with
an heterogeneous anisotropic medium
(for instance deduced from a geometric transform),
the Laplace problem reads as:
\begin{equation}
- \nabla\cdot ({\bf A}\nabla u) = f \hbox{ dans } D \; , \; u=0 \hbox{ on } \Sigma_1 \hbox{ and } \Sigma_2 \; ,
\end{equation}
where ${\bf A}$ is a matrix-valued function. The solution $u$ is now very different from above,
and its behaviour around the tip of the sector/conical shaped region therefore changes
dramatically. This simple analysis exemplifies the role of the transformed medium in the
regularization of the field scattered by a conical shaped bump surrounded by a carpet.

\subsection{Launching a SPP on a cylinder}
The goal of this section is to show that we can also launch a SPP on a cylindrical surface using the transformational optics
approach in order to guide smoothly the SPP which propagates on a flat metal surface to the top of a cylindrical bump
lying on the surface.

\begin{figure}[h]
\begin{center}
\resizebox*{10cm}{!}{\includegraphics{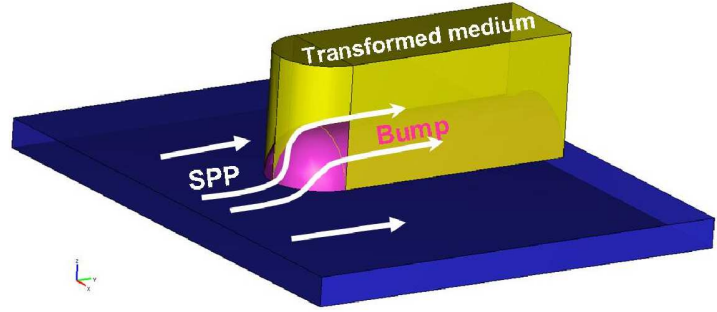}}
\caption{Principle of a hybrid tubular carpet for SPP. The transformed medium consists of two parts: The pink part
amounts to projecting a semi-hemisphere on a flat surface, while the yellow part amounts to projecting one half of circular cylinder on a flat surface. The trajectory of the SPP is shown with white arrows for illustrative purpose.}
\end{center}
\label{fig4}
\end{figure}

More precisely, the transformed medium consists of two parts:
On one hand, the quarter of a sphere (shown in pink color in figure \ref{fig4}) requires the following transform:
\begin{equation}
z_1 = \sqrt{0.04-x^2-(0.22+y)^2} \quad ; \quad z_2 = 0.5 
\end{equation}

On the other hand, the cylindrical part  (shown in yellow color in figure \ref{fig4}) requires the following transform:
\begin{equation}
z_1 = \sqrt{0.04-x^2} \quad ; \quad z_2 = 0.5 
\end{equation}

\begin{figure}[h]
\begin{center}
\resizebox*{10cm}{!}{\includegraphics{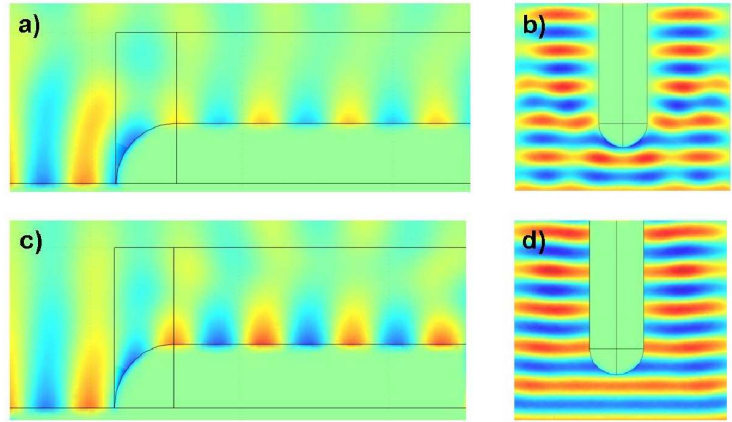}}
\caption{
SPP incident from the left on the hybrid tubular bump of Figure \ref{fig4} which consists of one half of hemisphere and one half of circular cylinder lying on a metal surface. Tubular bump on its own (upper panel): (a) Slice of magnetic field along (x,z) i.e. side view; (b) Slice of magnetic field along (x,y) i.e. top view;  Tubular bump with carpet (lower panel): (c) Slice of magnetic field along (x,z) i.e. side view; (d) Slice of magnetic field along (x,y) i.e. top view.}
\end{center}
\label{fig5}
\end{figure}

We depict in Figure \ref{fig4} the typical trajectory for a SPP launched on the tubular surface
surrounded by the carpet consisting of these two parts. We report in Figure \ref{fig5} the associated finite element
computations.

\subsection{Rotator}

In this section we consider an arbitrary shaped rotator described as follows:
\begin{equation}
R_1(\theta) = 0.4 \; R \; (1+0.2 \sin(3\theta)) \quad ; \quad R_2(\theta) = R \; (1+0.2 \sin(3\theta)+0.2 \cos(4\theta))
\quad ; \quad R = 0.4 \; .
\end{equation}
One can see in Figure \ref{fig7}(b) that the SPP is smoothly rotated within the rotator. Moreover, the concentrator is
itself invisible for SPPs as the wavefronts are unperturbed outside this metamaterial.

\begin{figure}[h]
\begin{center}
\resizebox*{10cm}{!}{\includegraphics{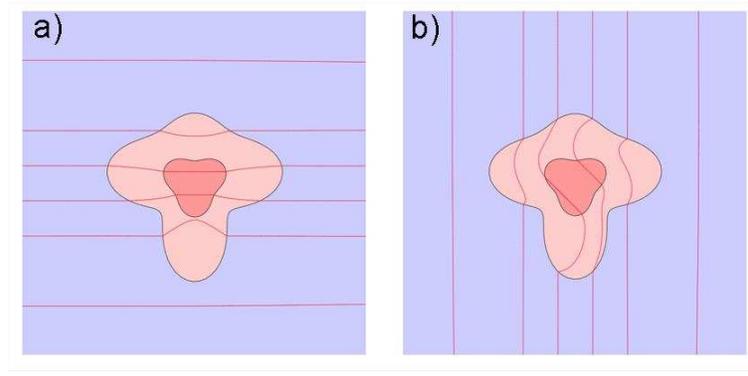}}
\caption{Principle of SPP scattering by a concentrator and a rotator: (a) The wavefront of an SPP incident from the top on a concentrator is squeezed within the meta-surface, while going unperturbed elsewhere;
(b) The wavefront of an SPP incident from the left on a rotator is rotated within the meta-surface, while going unperturbed elsewhere. We note that in both systems, the wavefronts are parallel in the inner
core, and are distorted inside the coating.}
\end{center}
\label{fig6}
\end{figure}


More precisely, we consider the geometric transform:
\begin{equation}
\left\{ \begin{array}{c c l}
            r' & = & r \\
	    \theta' & = & \alpha.r + \beta \\
	    z' & = & z\\
           \end{array} \right.\quad  \text{with}
\quad \alpha = \dfrac{\theta_o}{R_1(\theta)-R_2(\theta)} \quad ;
\quad \beta = \theta+\dfrac{R_2(\theta).\theta_o}{R_2(\theta)-R_1(\theta)}
\end{equation}\\
\bigskip
\begin{equation}
{\bf J}_{rr'} = \left( \begin{array}{c c c}
            1 & 0 & 0 \\
	    -\alpha & c_{22} & 0 \\
	    0 & 0 & 1 \\
           \end{array} \right) \quad \Longrightarrow
\quad {\bf T}^{-1} =  R(\theta) \left( \begin{array}{c c c}
            c_{22} & \alpha.r & 0 \\
	    \alpha.r & \dfrac{1+\alpha^2.r^2}{c_{22}} & 0 \\
	    0 & 0 & c_{22}\\
           \end{array} \right) R(\theta)^{T}
\end{equation}\\

\begin{multline}
c_{22} = \dfrac{\partial \theta}{\partial \theta'} =
1 - \dfrac{\theta_o.r}{\left( R_2(\theta)-R_1(\theta) \right)^2}.\left( \dfrac{\partial R_2(\theta)}{\partial \theta'} 
- \dfrac{\partial R_1(\theta)}{\partial \theta'}\right) \\
- \dfrac{\theta_o}{\left( R_2(\theta)-R_1(\theta) \right)^2}
\left(R_2(\theta) \dfrac{\partial R_1(\theta)}{\partial \theta'} -
R_1(\theta) \dfrac{\partial R_2(\theta)}{\partial \theta'}\right)
\end{multline}\\

\subsection{Concentrator}
\begin{figure}[h]
\begin{center}
\subfigure[Arbitrarily shaped concentrator]{
\resizebox*{5cm}{!}{\includegraphics{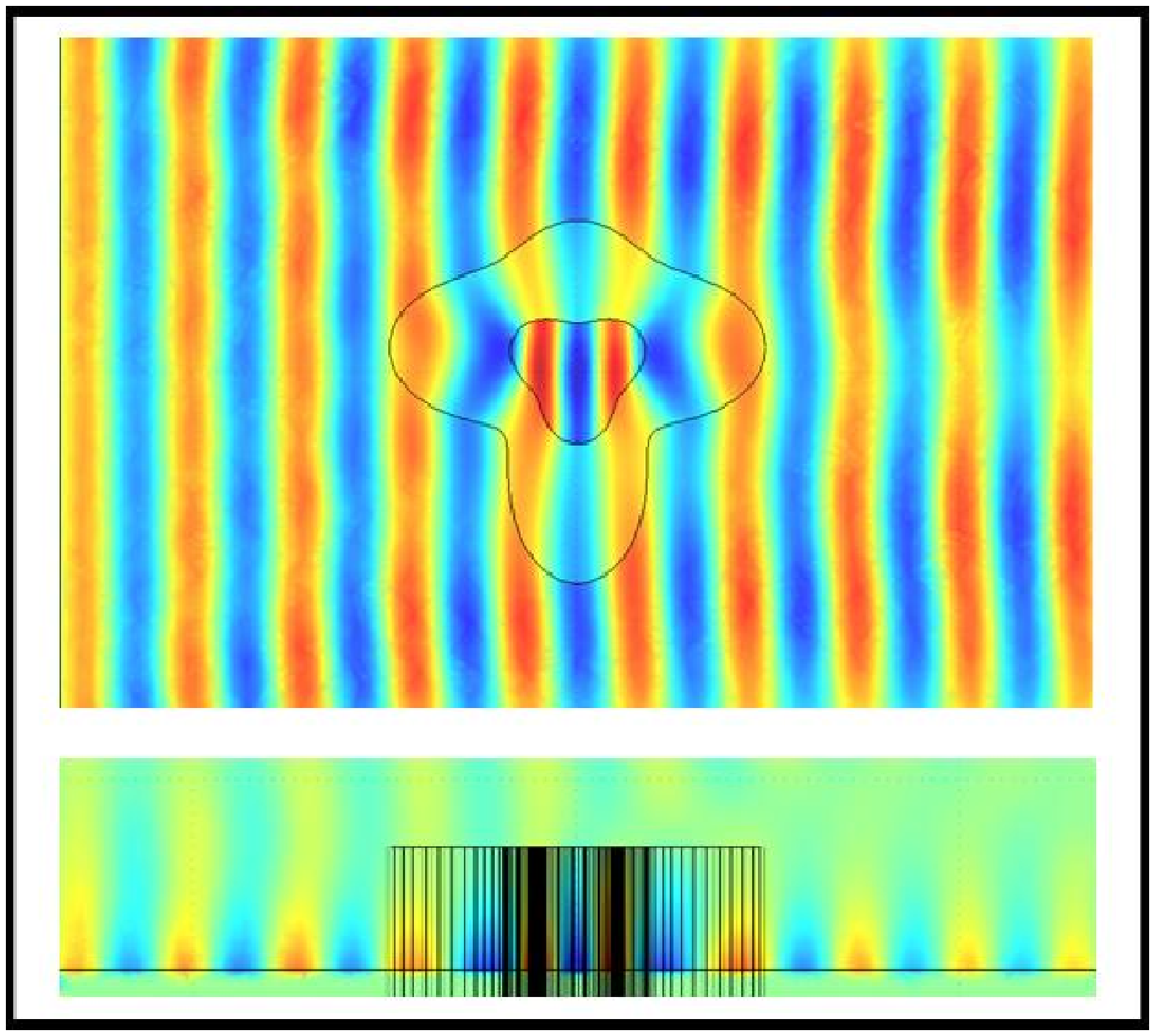}}}
\subfigure[Arbitrarily shaped rotator ]{
\resizebox*{5cm}{!}{\includegraphics{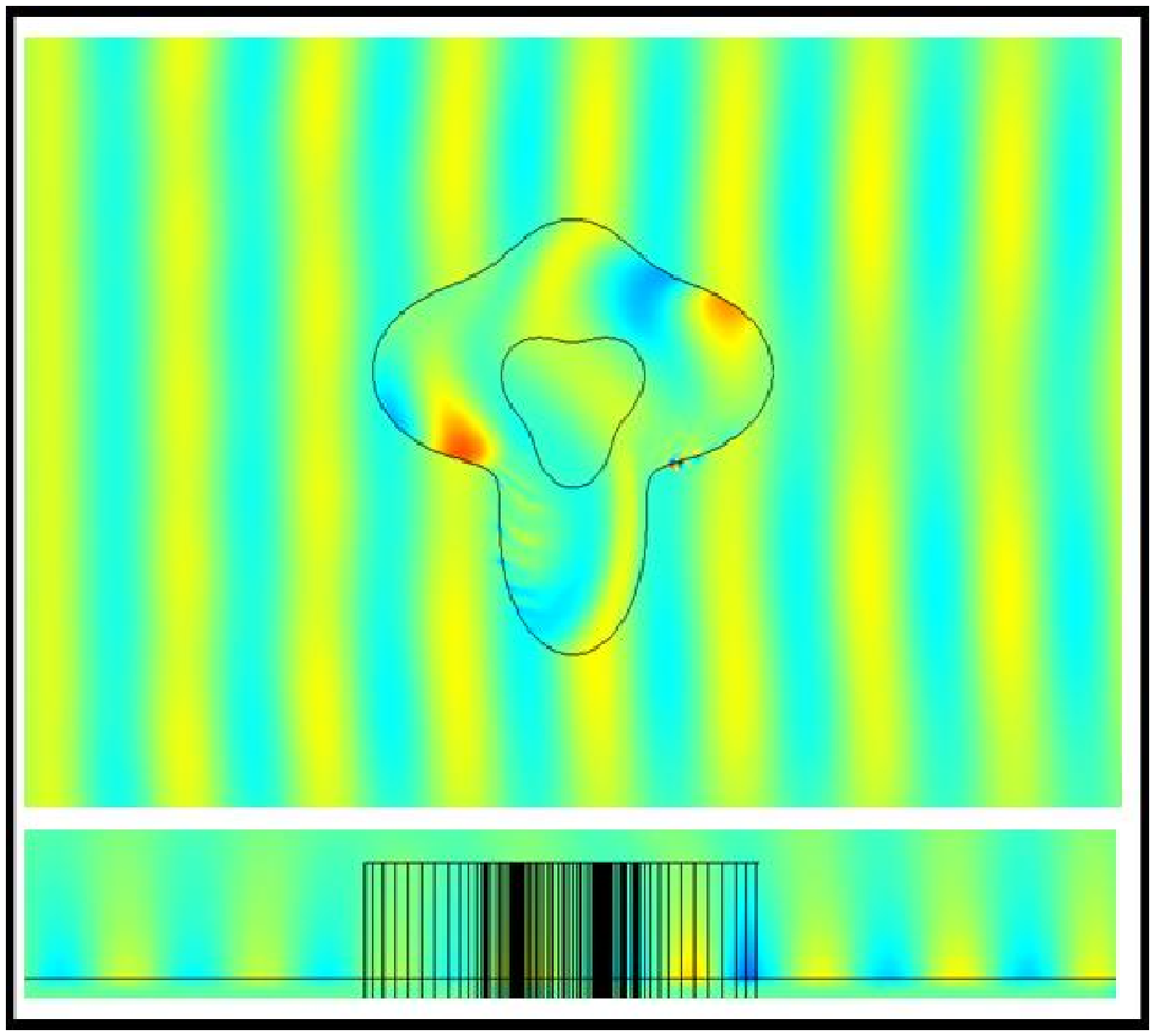}}}
\caption{Full-wave simulations for a concentrator and rotator: Phase representation of the y-polarized magnetic field for a SPP
incident from the left (x-axis) with top view along (x,y) and side view along (x,z). (a) Concentrator; (b) Rotator.}
\label{sample-figure}
\end{center}
\label{fig7}
\end{figure}

In this section we consider an arbitrary shaped concentrator described as follows:
\begin{equation}
\left\{ \begin{array}{ccl}
           R_1(\theta) = 0.4 R (1+0.2 \sin(3\theta))\quad ; \quad R_2(\theta) = 0.6 R (1+0.2 \sin(3\theta)) \\
           R = 0.4 \quad ; \quad R_3(\theta) = R(1+0.2(\sin(3\theta) + \cos(4\theta))) \;  
          \end{array} \right.
\end{equation}
One can see in Figure \ref{fig7}(a)
that the SPP is smoothly squeezed within the concentrator. Moreover, the concentrator is
itself invisible for SPPs as the wavefronts are unperturbed outside this metamaterial.

More precisely, we consider the following geometric transform:
\begin{equation}
\left\{ \begin{array}{c c l}
            r' & = & \alpha.r + \beta\\
	    \theta' & = & \theta \\
	    z' & = & z\\
           \end{array} \right.\quad 
\text{with} \left\{ \begin{array}{c c l}
        \alpha = \dfrac{R_1(\theta)}{R_2(\theta)} & \beta = 0 & (0 \leq r \leq R_1(\theta))\\
	& & \\
	\alpha = \dfrac{R_3(\theta)-R_1(\theta)}{R_3(\theta)-R_2(\theta)} &
 \beta = R_3(\theta)\dfrac{R_1(\theta)-R_2(\theta)}{R_3(\theta)-R_2(\theta)} & (R_1(\theta) \leq r \leq R_3(\theta))\\
        \end{array}\right.
\end{equation}

\begin{equation}
{\bf J}_{rr'} = \left( \begin{array}{c c c}
            1 & 0 & 0 \\
	    -\alpha & c_{22} & 0 \\
	    0 & 0 & 1 \\
           \end{array} \right) \quad \Longrightarrow
\quad {\bf T}^{-1} =  R(\theta) \left( \begin{array}{c c c}
            \dfrac{(r-\beta)^2+c_{22}^2.\alpha^2}{(r-\beta).r} & -\dfrac{c_{22}.\alpha}{r-\beta} & 0 \\
	    -\dfrac{c_{22}.\alpha}{r-\beta} & \dfrac{r}{r-\beta} & 0 \\
	    0 & 0 & \dfrac{r-\beta}{\alpha^2.r}\\
           \end{array} \right) R(\theta)^{T}
\end{equation}

\textbullet \hspace{0.2cm} $0 \leq r \leq R_1(\theta)$ :

\begin{equation}
c_{22} = \dfrac{\partial r}{\partial \theta'} =
-\dfrac{r}{R_1(\theta)^2}\left(R_2(\theta)\dfrac{\partial R_1(\theta)}{\partial \theta}-
R_1(\theta)\dfrac{\partial R_2(\theta)}{\partial \theta}\right)
\end{equation}

\textbullet \hspace{0.2cm} $R_1(\theta) \leq r \leq R_3(\theta)$ :

\begin{multline}
c_{22} = \dfrac{-1}{\left( R_2(\theta)-R_1(\theta)\right)^2} \left\{
\dfrac{\partial R_1(\theta)}{\partial \theta}(R_3(\theta)-r)(R_3(\theta)-R_2(\theta))\right.\\
\left. + \dfrac{\partial R_2(\theta)}{\partial \theta}(R_1(\theta)-R_3(\theta))(R_3(\theta)-r) -
\dfrac{\partial R_3(\theta)}{\partial \theta}(R_2(\theta)-R_1(\theta))(R_1(\theta)-r) \right\}
\end{multline}

\section{Geometric potential for SPPs on a weakly curved interface with an anisotropic medium}

\begin{figure}[h]
\begin{center}
\resizebox*{9cm}{!}{\includegraphics{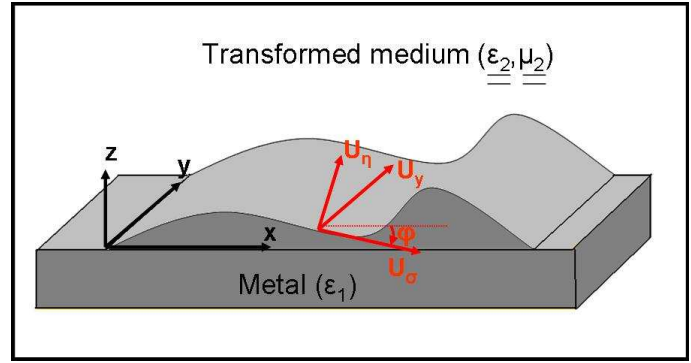}}
\caption{
Diagrammatic view of a metal-transformed medium interface and associated global and local coordinates systems.
}
\end{center}
\label{fig8}
\end{figure}

In order to illustrate analytically the control of the evanescent component of the SPP,
we consider the case of a curved surface which is invariant along the $x$-direction, see Figure \ref{fig8}.
Following the paper by Della Valle and Longhi \cite{tropdelavalle} where a detailed discussion of the
role played by the geometric potential  on the SPP envelope in the direction of propagation
led to the evolution equation:
\begin{equation}
-i{\lambda}\frac{\partial F}{\partial z}=
-\frac{\lambda^2}{4\pi n_{eff}}\frac{\partial^2 F}{\partial \sigma^2}
+ V_{Top}(\sigma) F
\end{equation}
in the case of two isotropic media.

Della Valle and Longhi used a two-scale asymptotic expansion in the electromagnetic field, assuming that the local curvature of
the interface is small, i.e. $\mid R \sim \alpha^{-2}\mid \gg 1$ with $R^{-1}=d\phi/d\sigma$ and $\alpha$ is a small positive
parameter.
Importanttly, the asymptotic expansions are assumed as follows:
\begin{equation}
{\bf u}=\sum_{i=0}^\infty\alpha^{2i}{\bf u}^{(i)} \; , \; {\bf v}=\sum_{i=0}^\infty\alpha^{2i+1}{\bf v}^{(2i+1)} 
\end{equation}
with ${\bf u}=(cB_\sigma,E_\eta,E_z)$ and ${\bf v}=(E_\sigma,cB_\eta,cB_z)$ i.e. for p- and s-field components. Moreover,
the corrective terms are functions of the slow variable $\sigma_1=\alpha\sigma$.

The former authors obtain a hierarchy of auxilliary equations at successive powers of $\alpha$. At the leading order $(\sim\alpha^0)$,
they retrieve the usual effective index of the SPP:
\begin{equation}
n_{eff}=\sqrt{\varepsilon_1\varepsilon_2/(\varepsilon_1+\varepsilon_2)}
\end{equation}
and proceed with the interface conditions for the field to obtain the asymptotic expansion of the decay length:
\begin{equation}
\gamma_{j}=\sum_{i=0}^\infty\gamma_i^{(j)} \; , \; j=1,2
\end{equation}  
where the corrective terms $\gamma_i^{(j)}$ undergo some oscillations on the slow spatial scale $\sigma_1$.

Regarding the penetration depth of the SPP in the 2 media, Della Valle and Longhi deduce the following expression:
\begin{equation}
\gamma_{1,2}^2=\frac{2 \varepsilon_{2,1}^2-\varepsilon_{1,2}^2 \varepsilon_1 \varepsilon_1}{2R{\varepsilon_1^2-\varepsilon_1^2}} \; ,
\end{equation}
from the solvability condition:
\begin{equation}
2n_{eff}\frac{\partial F}{\partial z_2}+\i\frac{\partial^2 F}{\partial\sigma_1^2}-\i\frac{n_{eff}^2}{R}\frac{\gamma}{\varepsilon}F=0 \; ,
\end{equation}
at order $\alpha^2$.

We performed a similar analysis for a transformed (anisotropic heterogeneous) medium above
the curved surface.

The transformed medium is assummed to
be described by diagonal tensors of permittivity and permeability
\begin{equation}
\underline{\underline{\varepsilon'}}(u,v,w)=\rm{diag}(\varepsilon_\sigma,\varepsilon_\eta,\varepsilon_z)
\; , \; \underline{\underline{\mu'}}(u,v,w)=\rm{diag}(\mu_\sigma,\mu_\eta,\mu_z) 
\end{equation}
where the geometric potential is given by:
\begin{equation}
V_{top}=\frac{\lambda n_{eff}}{2R}*\sqrt{-\frac{1}{\varepsilon_1+\varepsilon_2}}
\end{equation}

In the local coordinate basis, we obtain the same evolution equation, with a new effective index
\begin{equation}
n_{eff}=\sqrt{\frac{\varepsilon_{z2}\varepsilon_1\mu_{\eta 2}\varepsilon_{1}-\varepsilon_{\sigma_2}}{\varepsilon_1^2-
\varepsilon_{\eta 2}\varepsilon_{\sigma2}}}
\end{equation}
where the geometric potential is now given by:
\begin{equation}
V_{top}=\frac{\lambda n_{eff}}{2R}*\sqrt{\frac{(\varepsilon_1\mu_{yy2}-\varepsilon_{xx2})}{\varepsilon_1^2-\varepsilon_{xx2}\varepsilon_{zz2}}}
\end{equation}

\section{Concluding remarks and perspectives}

We have demonstrated the full control of surface plasmon polariton propagation
by extending the transformational optics tools to the area of plasmonics.  We have shown that we can
markedly reduce the scattering of SPP on a bumped surface thanks to invisibility carpets whose design
is the same as for an out-of-plane electromagnetic field incident upon the metal surface. The versatility
of the designs proposed in this paper illustrates the power of transformational plasmonics. 

\subsection{Acknowledgements}

M. K. and G. D. are thankful for the PhD scholarship from the French Ministry of Higher Education and Research,
and the University of Aix-Marseille III.

\label{lastpage}

\end{document}